\documentclass[intlimits,twoside,a4paper]{article}

\usepackage[cp1251]{inputenc}

\usepackage[eqsecnum]{cmpj3}

\usepackage{bm}

\issue{2018}{21}{3}{33602}
\doinumber{10.5488/CMP.21.33602}

\title[A machine learning approach to the Berezinskii-Kosterlitz-Thouless transition]{A machine learning approach to the Berezinskii-Kosterlitz-Thouless transition in classical and quantum models}
\author[M. Richter-Laskowska, H. Khan, N. Trivedi, M.M. Ma\'ska]{M. Richter-Laskowska\refaddr{US}, H. Khan\refaddr{OH}, N. Trivedi\refaddr{OH}, M.M. Ma\'ska\refaddr{US}}
\addresses{
\addr{US}Institute of Physics, University of Silesia, 75 Pu\l ku Piechoty 1, 41-500 Chorz\'ow, 
Poland
\addr{OH}Department of Physics, The Ohio State University, 191 W. Woodruff Ave., Columbus, Ohio 43210, USA
}

\date{Received May 31, 2018, in final form August 16, 2018}

\begin{document}

\maketitle
\begin{abstract}
The Berezinskii-Kosterlitz-Thouless transition is a very specific phase transition where all thermodynamic quantities are smooth. Therefore, it is difficult to determine the critical temperature in a precise way. In this paper we demonstrate how neural networks can be used to perform this task. In particular, we study how the accuracy of the transition identification depends on the way the neural networks are trained. We apply our approach to three different systems: (i) the classical XY model, (ii) the phase-fermion model, where classical and quantum degrees of freedom are coupled and (iii) the quantum XY model.
\pacs 64.60.-i, 05.70.Fh, 07.05.Mh
\end{abstract}

\section{Introduction}
In many cases, thermodynamic phase transitions are clearly visible with well defined and easily identifiable critical points. In the Landau picture, we define a macroscopic order parameter that is non-zero in the ordered phase and vanishes when we cross the critical temperature. Typically, the transition is signaled by a discontinuity or divergence of some thermodynamic quantities such as specific heat or magnetic susceptibility.
There also exist unconventional phase transitions which are much more difficult to find. One example are topological phase transitions, connected to the formation of topological defects, such as dislocations in two-dimensional crystals, vortices in two-dimensional superconductors and so on. The proliferation of these defects leads to the Berezinskii-Kosterlitz-Thouless (BKT) phase transition \cite{BKT1,BKT1a,BKT2,BKT2a} where thermodynamic quantities behave smoothly.
Therefore, traditional detection methods based on, e.g., the divergence of the specific heat or the spin susceptibility, cannot be used. In numerical approaches, the main difficulty with the identification of the
BKT transition stems from the fact that the interaction between the topological charges depends logarithmically on the spatial separation. Therefore, numerical results converge very slowly with the size of the system, and a precise determination of the critical temperature is a computationally challenging task.
The standard approach is based on the scaling properties of, e.g., the spin stiffness or superfluid density. This is particularly difficult for quantum models, where numerical methods are usually very involved and memory- and time-consuming.

Besides the traditional methods which rely on the idea of the order parameter or quantities accessible in experiments, such as the specific heat or magnetic susceptibility, attempts to use artificial neutral networks to identify phases in condensed matter and transition between them have been recently undertaken \cite{phtrans,ising,ising1,phtrans1,phtrans2,phtrans3,phtrans4,phtrans5,phtrans6,phtrans7,
qml,qml1,melko2018,sarma,zhang2018,Nieva}. These new methods have proven to be accurate and reliable for classical Ising-type models \cite{phtrans,ising,ising1}. Less spectacular progress has been made for quantum systems \cite{qml,qml1} and for systems with topological phase transitions \cite{melko2018, sarma, zhang2018, Nieva}. In the latter case, the main difficulty comes from the fact that the
 topologically ordered phase is not described by a local order parameter 
 \cite{BKT1,BKT1a,BKT2,BKT2a}. Instead, its 
 formation is connected with suppression of non-local topological defects which are difficult 
 to identify.  

In this paper, we demonstrate the application of machine learning approaches 
to identify topological transitions in a few different types of two-dimensional classical and 
quantum systems.
In particular, we study the classical XY (c-XY) model, the phase-fermion (PF) 
\cite{pf} where the interaction is only between quantum and classical degrees of
freedom, and the fully quantum XY (q-XY) model. 

The first of these models has already been thoroughly analyzed in  \cite{melko2018}.
The authors have shown there that treating spin configurations as raw images in the case of a feed-forward 
network does not lead to the correct value of the critical temperature. Instead, they propose to
preprocess the spin configurations into vorticity and then use the results to train two kinds 
of artificial neural networks (ANN): a one-layer feed-forward network and a deep convolutional network. In both cases, the results are scaled 
with the system size towards the correct value of the critical temperature, but the one-layer network
performed poorly for large systems. In the present approach, we do not have convolutional layers, but 
we use a deep feed-forward network composed of four fully-connected layers (we learned that the choice of particular meta-parameters is not crucial for the network performance). What is important is that, instead of 
using raw spin configurations where each spin is represented by a number from 0 to $2\piup$ (which 
gives rather inaccurate results, as demonstrated in  \cite{melko2018}), we represent the configurations
as vectors of sines and cosines of the spin angles, which reflects the system's symmetry.

\section{Models}

The first model describes classical spins of unit length in a square lattice with
a nearest neighbor interaction given by the following Hamiltonian
\begin{equation}
    H_\text{c-XY}=-J\sum_{\langle i,j\rangle}\cos\left(\theta_i - \theta_j\right),
\end{equation}
where $J$ is the coupling between the spins, and $\theta_i$ describes the direction of spin $i$. It is known that this model exhibits the BKT phase transition, and precise 
finite-size scaling gives the critical temperature $T_{\rm BKT}\approx 0.8935$ in units of 
$J$ \cite{sandvik}.

In the next model, classical spins $\theta_i$ interact with fermions. The Hamiltonian 
of the PF model is given by
\begin{equation}
    H_{\rm PF}=-t\sum_{\langle i,j\rangle,\sigma}\hat{c}^\dagger_{i\sigma}\hat{c}_{j\sigma}
+g\sum_i\left(\re^{\ri\theta_i}\hat{c}_{i\uparrow}\hat{c}_{i\downarrow}+{\rm h.c.}\right),
\label{PF}
\end{equation}
where $\hat{c}^\dagger_{i\sigma}\ \left(\hat{c}_{i\sigma}\right)$ is an operator
that creates (annihilates) spin-$\sigma$ electron at the lattice site $i$, and $g$
describes the strength of the interaction between the classical and quantum degrees of 
freedom. We set the hopping integral $t$ as the energy unit ($t=1$). 
In this model, 
the fermions mediate an effective interaction between the
classical spins $\theta_i$ which also leads to the BKT phase transition. The critical temperature is a function of $g$, and for $g=2$ Monte Carlo (MC) simulations give
$T_{\rm BKT}\approx 0.12$ \cite{pf}. The PF model can be treated as an approximation of the 
boson-fermion model \cite{bf}, valid when fluctuations of the number of bosons 
at a lattice site can be neglected.

The Hamiltonian of the last model, the quantum XY model, is given by
\begin{equation}
    H_\text{q-XY}=\frac{E_\text {c}}{2}\sum_i\hat{n}_i^2-J\sum_{\langle 
    i,j\rangle}\cos\left(\hat{\theta}_i - \hat{\theta}_j\right),
\end{equation}
where $\hat{n}_i$ is the number operator that is canonically conjugate to
the quantum phase operator $\hat{\theta}_i$, and $E_\text {c}$ is the charging energy.

To train neural networks and to classify phases, one needs extensive sets 
of spin configurations generated at different temperatures. They were produced 
with the help of the MC simulations. In the case of the classical XY model, 
we directly used the Metropolis algorithm. For the PF model, we also used the
Metropolis algorithm, but in each MC step (i.e., for each generated spin configuration) we 
needed to diagonalize the fermionic
Hamiltonian (\ref{PF}) \cite{FK}. For the quantum XY model, we used a quantum MC method in 
which the Hamiltonian was mapped to a classical action of spins on an effective (2+1)D 
lattice. We then used a Wolff cluster algorithm to simulate these spins. In all
cases, the simulations were performed on $16\times 16$ systems.

\begin{figure}[!b]
    \centering
    \includegraphics[width = 0.47\textwidth]{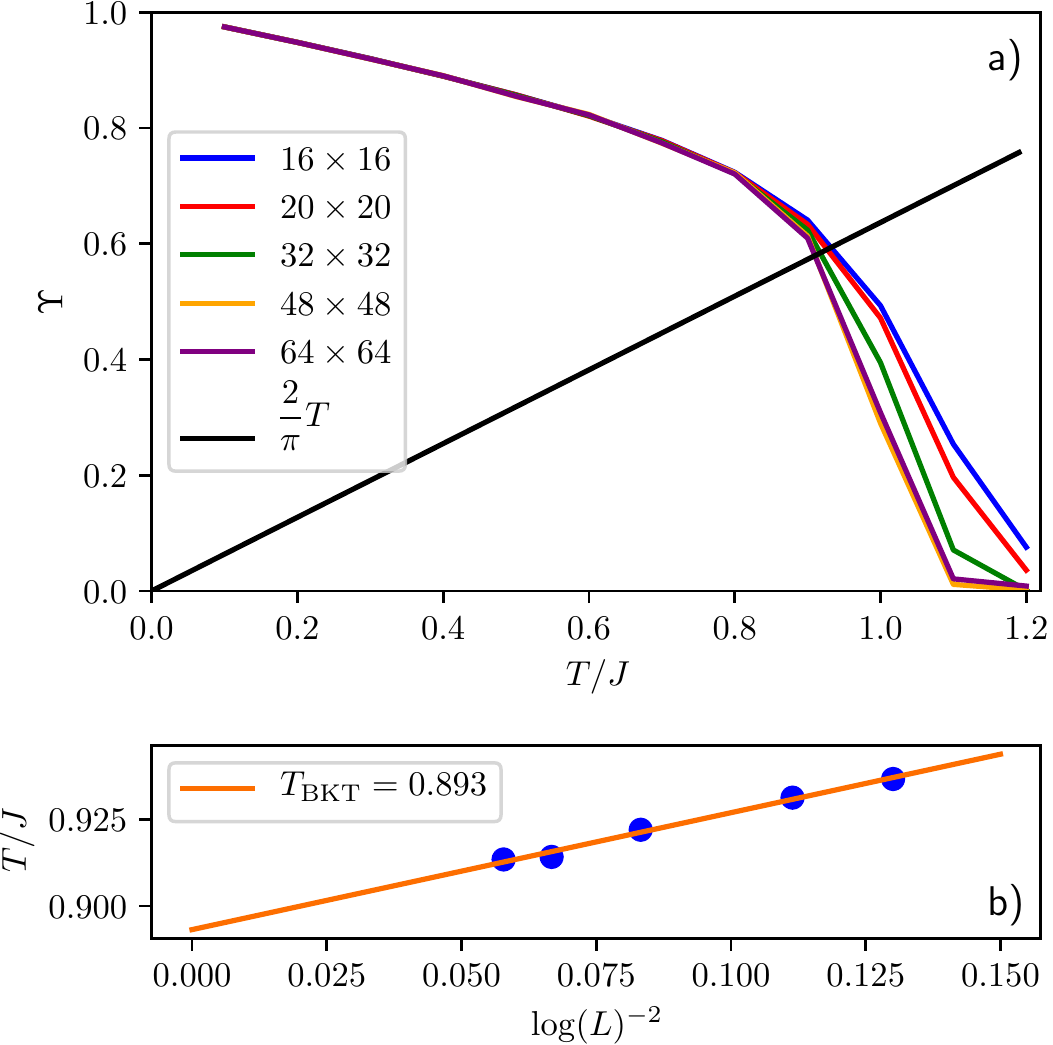}\qquad
    \includegraphics[width = 0.47\textwidth]{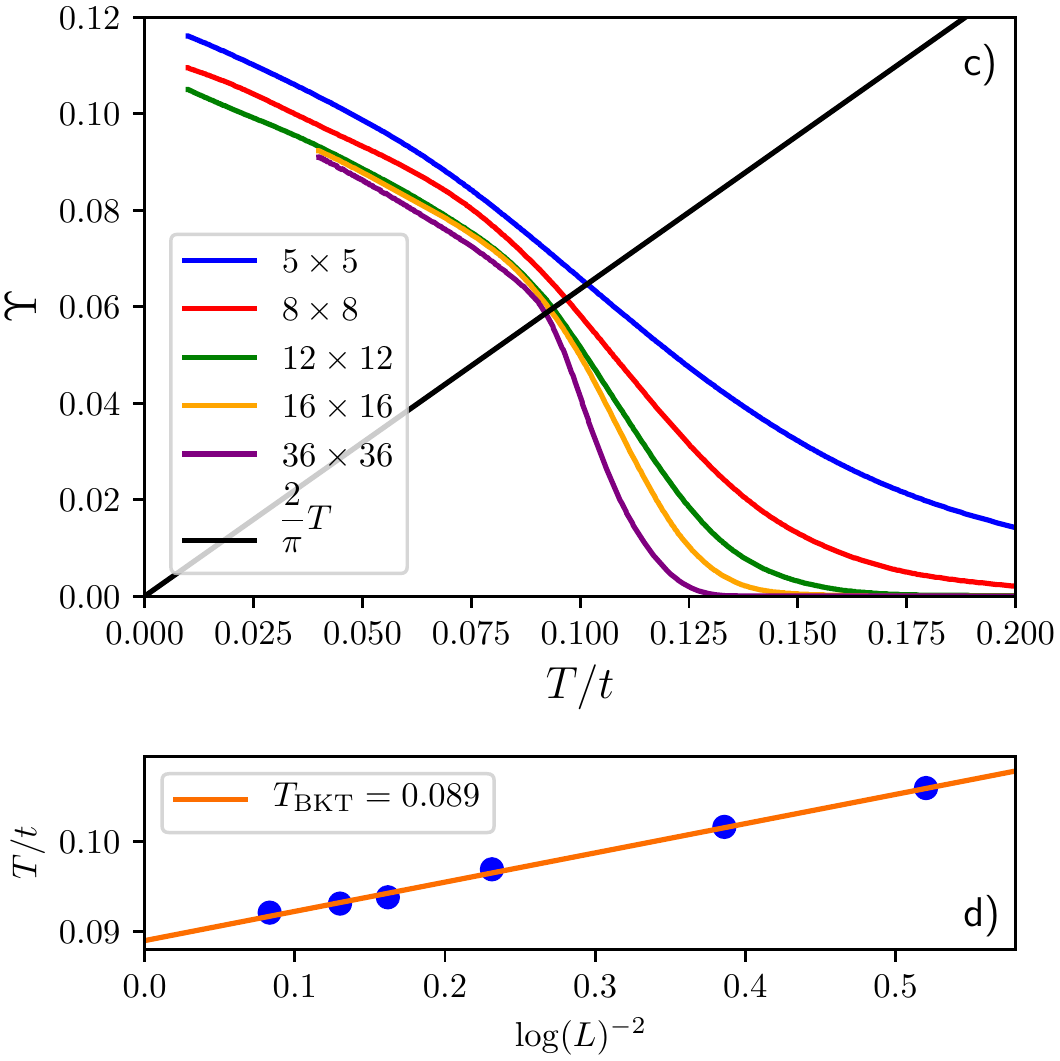}\vskip 5mm
    \includegraphics[width = 0.47\textwidth]{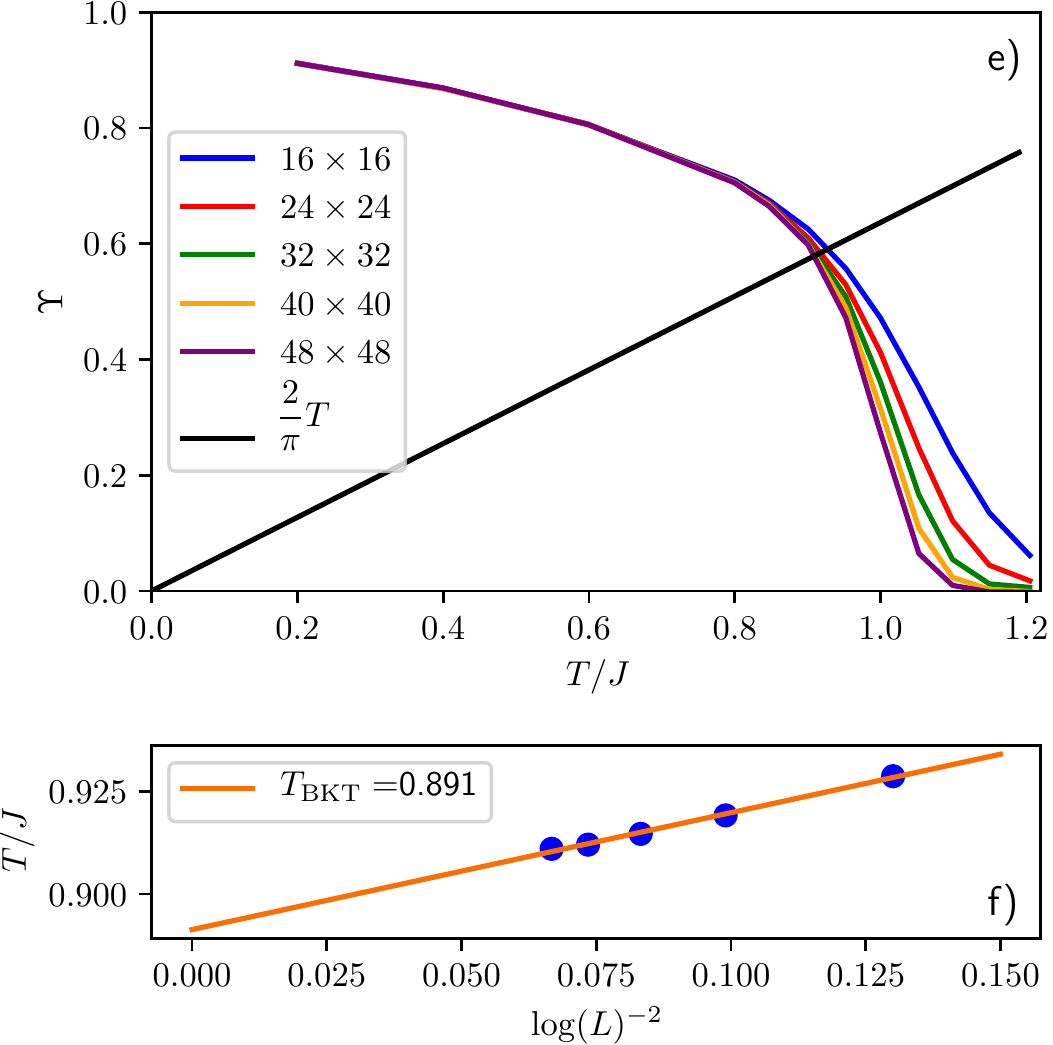}
    \caption{(Colour online) Helicity modulus for the c-XY model (a), PF model for $g=4t$ (c), and q-XY model for $E_\text {c}=0.1J$ (e). The corresponding critical temperatures extrapolated to 
    the thermodynamic limit are presented in panels (b), (d), and (f).
    Since $T^*(L)-T_{\rm BKT}\propto \log (L)^{-2}$, where $T^*(L)$ is
    the temperature at which $\Upsilon$ drops the most rapidly in a $L\times L$ system
    \cite{Schultka, Tomita, sandvik},
    the temperature is presented in these panels as a function of $\log (L)^{-2}$.
    }
    \label{fig:hel_mod}
\end{figure}

\begin{figure}[!t]
    \centering
    \includegraphics[height = 6.3cm]{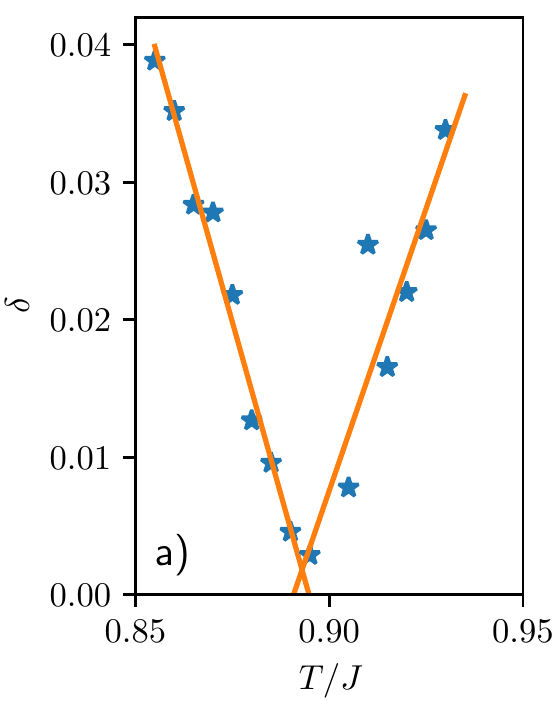}
    \includegraphics[height = 6.3cm]{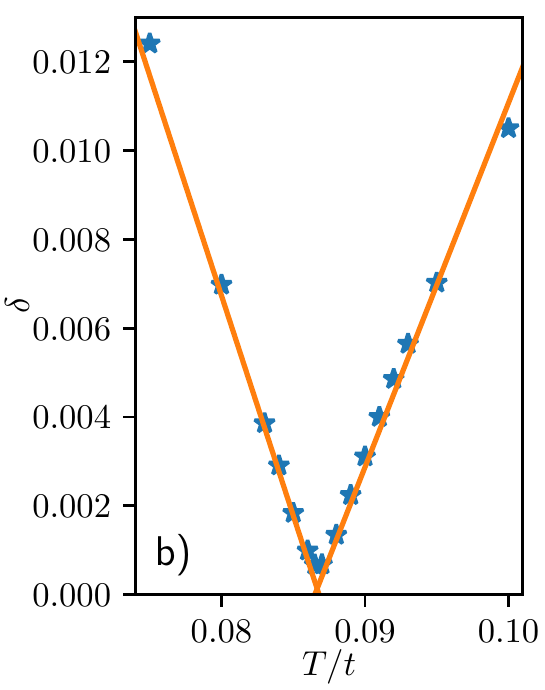}
    \includegraphics[height = 6.3cm]{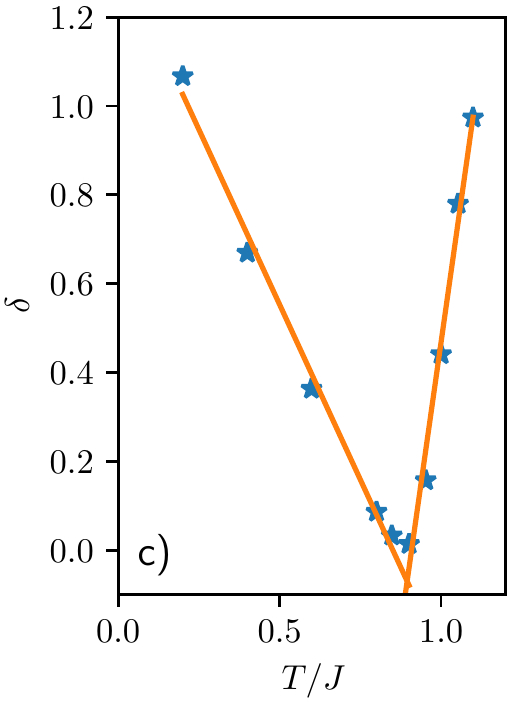}
    \caption{(Colour online) The root-mean-square error $\delta$ for fitting MC results 
    for the c-XY model (a), the PF model (b), and the q-XY model (c)
    to the RG predictions given by equation~(\ref{eq:RG}). 
    The sharp minimum of the fitting errors indicates $T_{\rm BKT}$. 
    The model parameters are the same as in figure \ref{fig:hel_mod}. 
    }
    \label{fig:fitting_errors}
\end{figure}

In all these models, the helicity modulus $\Upsilon$, at finite temperature defined as
the second derivative of the free energy with respect to an externally imposed global
twist across the sample \cite{helicity}, has a universal jump at the BKT transition. 
In the thermodynamic limit, it drops at $T_{\rm BKT}$ from 
$\frac{2}{\piup} T_{\rm BKT}$ to zero. However, in finite systems it evolves smoothly 
and converges only logarithmically to the thermodynamic limit, as shown in figures~\ref{fig:hel_mod}~(a), \ref{fig:hel_mod}~(c), and \ref{fig:hel_mod}~(e). 
A finite size scaling of the temperature at which
$\Upsilon(T)$ crosses $\frac{2}{\piup}T$ can give a rough estimate of $T_{\rm BKT}$.
This is demonstrated in figures~\ref{fig:hel_mod}~(b), \ref{fig:hel_mod}~(d), and \ref{fig:hel_mod}~(f).
It is also possible to determine $T_{\rm BKT}$ in a more precise way.
At the critical temperature, the helicity modulus
scales with the system size according to the Kosterlitz renormalization group (RG) equations
\cite{KRG},\\
\begin{equation}
    \Upsilon_L=\Upsilon_{L\to\infty}\left[1+\frac{1}{2}\frac{1}{\ln(L) + C}\right],
    \label{eq:RG}
\end{equation}
where $L$ is a linear size of the system and $C$ is a constant.
Therefore, if one fits $\Upsilon_L(T)$ calculated in MC simulations for different system sizes 
$L$ to the RG 
predictions, the fitting errors drop almost to zero exactly at $T_{\rm BKT}$ \cite{Weber}. 
In figure \ref{fig:fitting_errors} one can see the procedure.

\section{Artificial neural network}

Here, however, 
we study how the critical temperature can be determined with the help of ANN trained to 
identify the low- and high-temperature phases. When a system approaches the critical 
temperature, thermal fluctuations increase drastically so that the spin configurations can
be very different from configurations generated at very low temperatures. However, as long as $T$ is below 
$T_{\rm BKT}$, the system is still in the same low-temperature phase. The problem with 
systems where the BKT transition takes place, such as the c-XY, PF or q-XY models, is that 
in the low-temperature phase, MC simulations on finite clusters show finite magnetization, 
which according to the Mermin-Wagner theorem cannot exist in the thermodynamic limit. Since 
the MC results are used to train the ANN, it is possible that the network will learn 
finite-size features of the spin configurations, such as magnetization. It was proposed in 
 \cite{melko2018} that
this difficulty can be overcome by adding 
a convolutional layer designed to identify topological defects in spin configurations.
Then, the network learns configurations of vortices and antivortices instead of raw spin
configurations.

Our aim here is to show how the features that can be used to identify phases depend on
the distance from the critical temperature $T_{\rm BKT}$. The standard ANN-based approach 
to the 
problem of finding a phase transition is to train the ANN to recognize features
of the low- and high-temperature phases. Within the scheme of {\em supervised 
learning}, one needs to feed the ANN with labelled spin configurations generated at low 
and high temperatures. The network is supposed to extract characteristic 
patterns and to learn how to classify configurations which were not used at the training
stage. Usually, this is an easy task for configurations generated at very low or at very 
high temperatures, since they are clearly distinctive. However, to precisely determine
the critical temperature, the ANN must distinguish configurations generated slightly 
below $T_{\rm BKT}$ and slightly above $T_{\rm BKT}$. Thermal fluctuations in this regime make 
those configurations very different from the fully ordered  low-temperature
configurations and from completely random high-temperature configurations. The question
then is, whether an ANN trained at extreme temperatures will be capable of classifying the 
phases close to $T_{\rm BKT}$? 
In other words, do the distinctive features
learned by the ANN at the extremes  balance each other out just right in the
evaluation process such that the correct $T_{\rm BKT}$ is predicted?
To answer this 
question we trained the ANN at different distances from $T_{\rm BKT}$ and checked how 
the distance $|T-T_{\rm BKT}|$ affects the accuracy of finding $T_{\rm BKT}$. To be precise, we 
used MC simulations
to generate sets of configurations $\{C_1\},\,\{C_2\},\,\ldots,\{C_N\}$ at 
temperatures $T_1,\,T_2,\,\ldots,\,T_N$ ($T_1<T_2<\ldots <T_N$) below and 
above $T_{\rm BKT}$. For the c-XY, PF, and q-XY models, the ranges of temperatures were
from $0.1J$ to $1.6J$, from $0.02t$ to $0.2t$, and from $0.1J$ to $1.5J$, respectively.
Then, we generated sets of 
configurations representing different low- (${\cal L}_m$) and high-temperature (${\cal H}_m$) ranges:
\begin{equation}
    {\cal L}_m=\bigcup\limits_{i=1}^{m}C_i\hspace{5mm} \mbox{and}\hspace{5mm} 
    {\cal H}_m=\bigcup\limits_{i=N-m+1}^{N}C_i\,,
    \label{eq:training_sets}
\end{equation}
where $m<N/2$\ \ is the number of temperatures in each range. 
We randomly removed from  ${\cal L}_m$ and ${\cal H}_m$ some number of configurations
to keep the cardinality of these sets fixed. In this way,
our study, on how the critical temperature predicted by the ANN depends 
on the range of temperatures used to train it, was not affected by a different
number of configurations for different ranges.
In order to connect $m$ with a temperature range used in the trainings, we introduce 
$\tau$ as a measure of the relative temperature range:
\begin{equation}
    \tau = \frac{T_{\rm high}-T_{\rm low}}{T_{\rm BKT}-T_{\rm low}}\,,
\end{equation}
where $T_{\rm low}$ and $T_{\rm high}$ are the lowest and the highest temperatures
lower than $T_{\rm BKT}$ which were used to train the ANN.

Configurations from ${\cal L}_m$ and ${\cal H}_m$ were mixed together and shuffled 
and then they were used to teach the ANN to identify the low- and high-temperature phases. 

The spin configurations generated in MC are stored as numbers $\theta_i$ from 0 to $2\piup$. 
However, in order to take into account the character of classical two-dimensional spins,
the configurations were rewritten as an array composed of cosines and sines: 
$\cos\theta_1,\ldots,\cos\theta_N,\sin\theta_1,\ldots,\sin\theta_N$, where $N=L\times L$
is the number of lattice sites. This is equivalent to a representation by complex numbers
and has the advantage that almost parallel spins are represented by close numbers, which 
is not the case for the original representation by the angles $\theta_i$.

\begin{figure}[!b]
    \centering
    \includegraphics[width=0.7\textwidth]{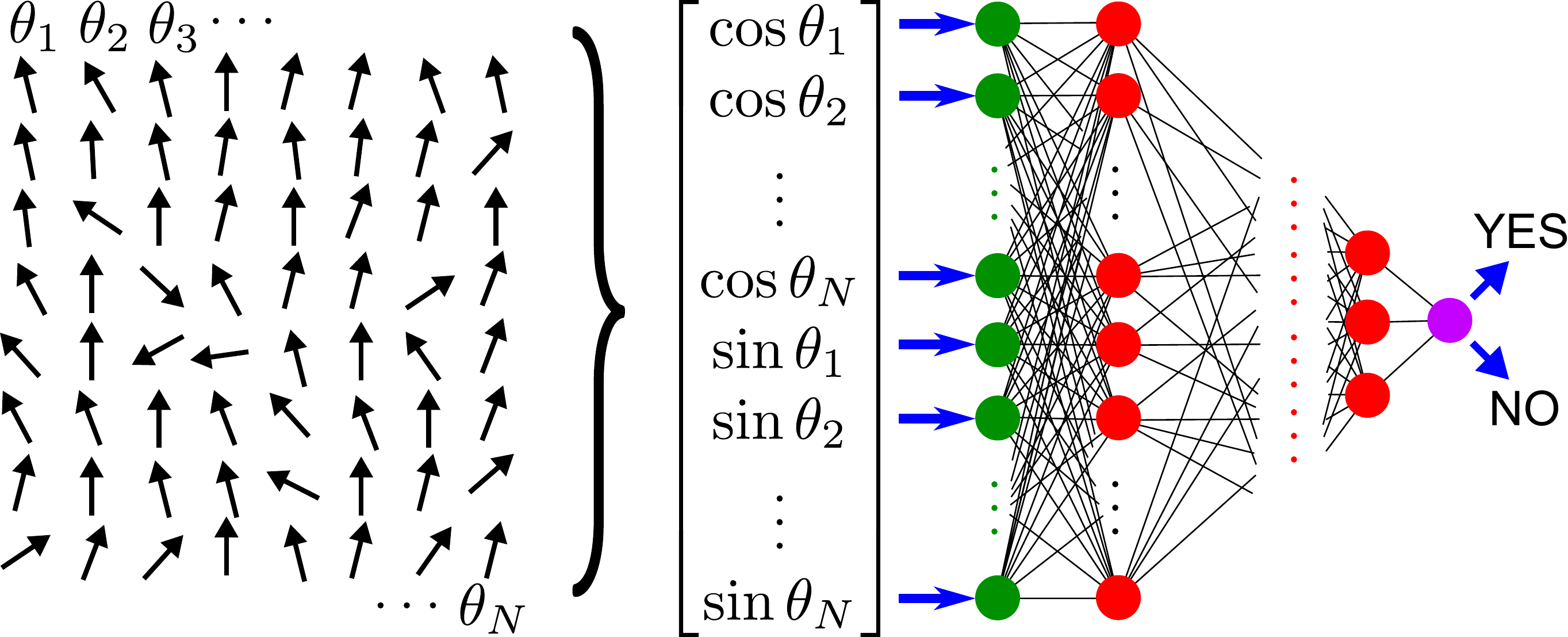}
    \caption{(Colour online) A scheme of a binary classification process in a feedforward
    ANN: a given spin configuration $\{\theta_i\},\ i=1,\ldots, N$, is 
    rewritten as a length-$2N$ vector
    $[\cos\theta_1,\ldots,\cos\theta_N,\sin\theta_1,\ldots,\sin\theta_N]$
    and is presented to the input layer of the ANN (green circles). Then, 
    activations flow across fully connected hidden layers (red circles)
    to the output layer composed of only one neuron (violet circle).}
    \label{fig:ann}
\end{figure}

Since the generation of extensive sets of spin configurations for different system sizes
and at different temperatures is a time-consuming task, especially for the PF and q-XY 
models, we used a technique known from 
machine learning-based image recognition to increase the number of configurations without
additional MC simulations. Namely, we transformed the original configurations according 
to symmetries of the system. We used periodic boundary conditions, which allowed us to apply 
translations to create new configurations. Other transformations were reflections and rotations.
Figure \ref{fig:ann} schematically shows how the ANN is used to classify spin configurations.

The ANN was implemented using the KERAS package with TENSORFLOW as the computational backend. We used 
a deep feedforward network with four fully connected hidden layers 
with 512, 192, 64, 16, and 16 neurons. Such a structure was a balance 
between the number of training epochs required for convergence and the time needed 
for one epoch. It turns out, however, that the metaparameters are not crucial for the
ANN performance. As the activation function,
a rectified linear unit (ReLU) was used in the hidden layers and sigmoid function in the 
output layer. The network was trained to minimize the distance 
between the MC data and the model predictions defined by the binary cross entropy.  
The loss function is given by 
\begin{equation}
    L=-\sum_i \left[ y_i \log p_i + (1-y_i)\log(1-p_i)\right],
\end{equation}
where $y_i$ are labels and $p_i$ are the corresponding predictions.

\begin{figure}[!t]
    \centering
    \includegraphics[width=0.65\textwidth]{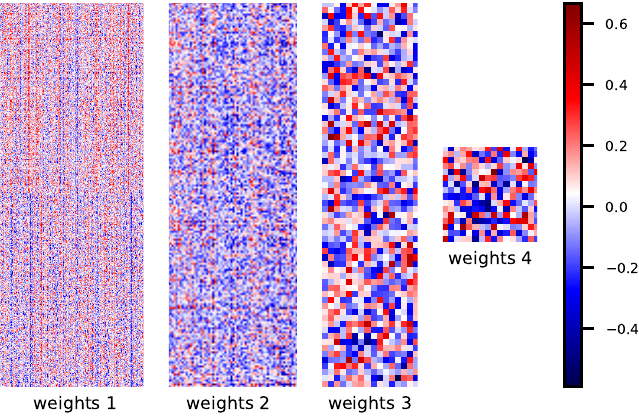}
    \caption{(Colour online) Visualization of the activations in a deep ANN composed of hidden layers of 512, 192, 
    64, 16, and 16 neurons. White colour corresponds to zero weight, blue to negative weight and red 
    to positive weight. ``Weights 1'' connect the output of layer 1 to layer 2, so they are represented by a rectangle with $512\times 192$ colour squares. Similarly, ``weights 2'' are 
    represented by a $192\times 64$ rectangle, etc.
    The network was initiated with all weights close to zero (we assumed a finite value
    $|w_i|<10^{-3}$ to avoid the {\em vanishing gradient problem} \cite{initialization}), so blue and 
    red parts indicate neurons which were activated during training.
    \label{fig:weights}}
\end{figure}

We found that with the Tikhonov regularization (L2) \cite{tikhonov}, the network 
performs better in the classification
of the low- and high-temperature phases. Figure \ref{fig:weights} shows an example of
learned weights used to classify phases of a $16\times 16$ system. One can see that despite a 
rather large size of the network, most of the neurons are activated. 

\section{Results}

\begin{figure}[!t]
    \centering
    \includegraphics[width=0.63\textwidth]{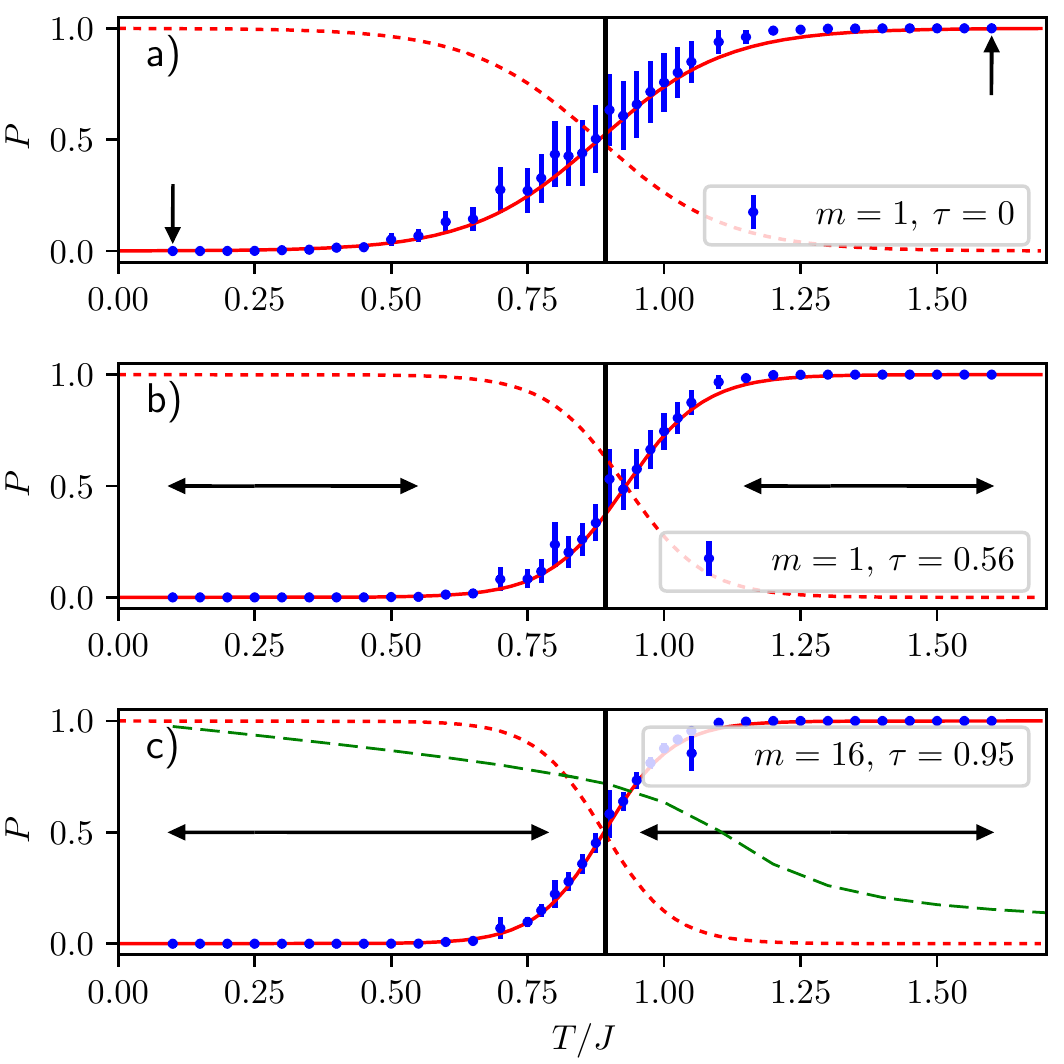}\raisebox{.3\height}{\includegraphics[width=0.35\textwidth]{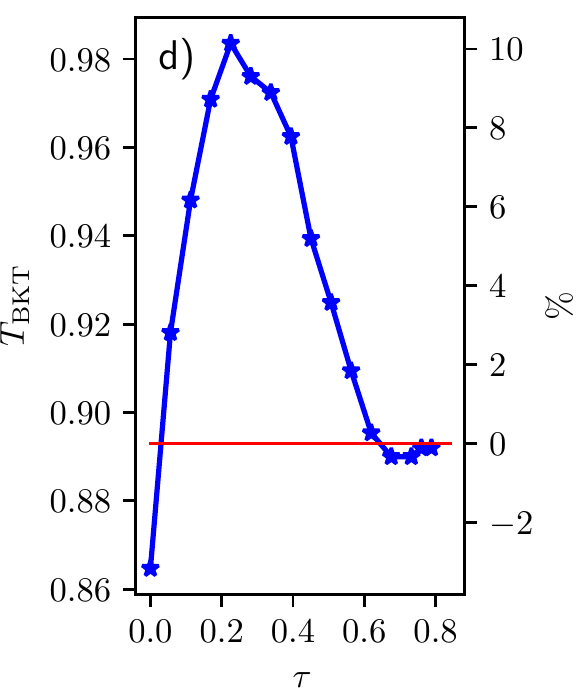}}
    \vspace{-2mm}
    \caption{(Colour online) Calculated by the ANN, the probability that a given configuration belongs to the 
    high-temperature phase of the c-XY model. The network was trained on labelled data generated
    according to equation (\ref{eq:training_sets}) for $m=1$ (a), $m=10$ (b), and $m=16$ (c). Then, 
    the network determined the probabilities 100 times for configurations generated at different
    temperatures each time starting from different 
    weights and biases. The vertical error bars show the standard deviation. The solid 
    red line is 
    the best fit $P(T)=0.5\tanh\left[\alpha\left(T-T_{\rm BKT}\right)\right]+0.5$,
    where $\alpha$ and $T_{\rm BKT}$ are fitting parameters. The dashed red line shows $1-P(T)$ which
    is the probability of being classified as a low-temperature phase. The black arrows indicate 
    the temperatures (a) or the ranges of temperatures (b), (c) used to train the network.
    The black vertical line indicates $T_{\rm BKT}$ determined from the RG equations.
    The dashed green line in panel (c) shows magnetization. Comparing panels (a)--(c) one
    can see that the for the c-XY model, the average critical temperature does not change 
    significantly with an increasing range of training temperatures, but the spread of
    the results shrinks. The temperatures at which the ANN was trained 
    and tested were (in units of $J$): from 0.10 to 0.70 and from 1.10 to 1.60 with 
    stepsize 0.05 and from 0.750 to 1.050 with stepsize 0.025.
    Panel (d) shows estimated $T_{\rm BKT}$ as a function of $\tau$. The horizontal red line 
    shows the critical temperature determined from fitting the MC results to the RG equation 
    (\ref{eq:RG}). The right-hand vertical axis shows 
    $(T_{\rm BKT}-T^0_{\rm BKT})/T^0_{\rm BKT}\times 100\%$, where $T_{\rm BKT}$ is the 
    average ANN prediction and $T^0_{\rm BKT}$ is the actual critical temperature.
    }
    \label{fig:prob_classXY}
\end{figure}

Figure \ref{fig:prob_classXY} shows how the predictions for $T_{\rm BKT}$ of the c-XY model 
calculated by the ANN depend on the way the network was trained.
In each case, we trained the neural network using a 10-fold cross validation technique 
\cite{cross-val} and repeated this procedure 10 times.
As a result, we obtained 100 possible values of probability $P$ that a given 
configuration belongs to the 
high-temperature phase (the probability that it belongs to the low-temperature phase is $1-P$).
Each time, the ANN was initialized with random weights and biases, and a larger spread of the 
predictions indicates a greater difficulty in an unambiguous classification of the phase.
The same method of multiple trainings starting from different random weights and biases was used in  \cite{phtrans6} to determine the standard error of the 
predicted critical temperature.
One can see in figure \ref{fig:prob_classXY} that even if the ANN was trained only at the 
extreme temperatures ($T=0.1$ and $T=1.6$,\ $m=1$) corresponding to a fully ordered and
completely random configurations, the average predicted $T_{\rm BKT}$ is not far from the 
actual value. This could be an accidental coincidence because with an increasing $m$
the deviation slightly increases, but always remains below 10\%, which can be seen in figure~\ref{fig:prob_classXY}~(d). 
For $m=16$, the deviation is smaller than the line width.
The problem, however, is that in this case, the uncertainty is large. So, for a 
precise determination of $T_{\rm BKT}$, the averaging over a large number of configurations 
generated at different temperatures is necessary. For example, for $m=1$,
the average prediction for 1000 statistically independent configurations is less than 1\% 
off the actual $T_{\rm BKT}$, 
but individual predictions are spread within the range of $\pm 18\%$ around this value.
This means that if one saves the computational
time required to generate configurations for training, more configurations should be 
generated for the evaluation stage. With an increasing width of the temperature range used for
training, the spread of the calculated probabilities decreases significantly. Figure~\ref{fig:prob_classXY}~(d)
shows how the average $T_{\rm BKT}$ depends on the number of different 
temperatures
used in training. On the right-hand vertical axis, showing the relative error, one can see that 
the error is always below 10\%, and with an increasing number $m$, it converges to the actual 
value of $T_{\rm BKT}$ for the c-XY model.

\begin{figure}[!t]
    \centering
    \includegraphics[width=0.63\textwidth]{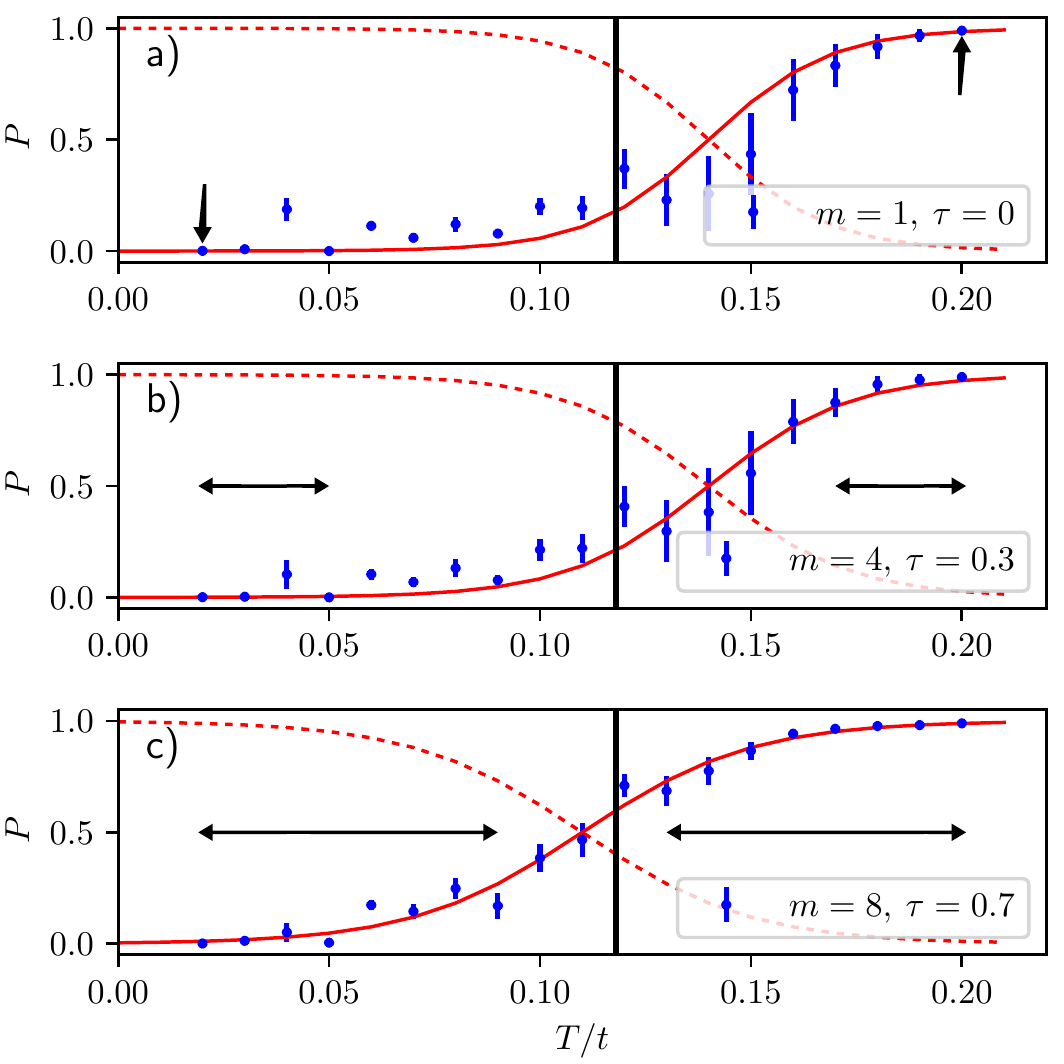}\raisebox{.3\height}{\includegraphics[width=0.35\textwidth]{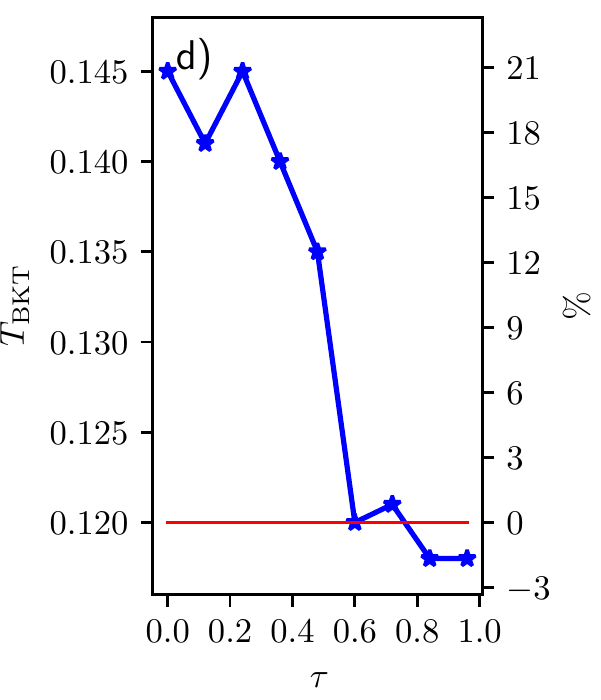}}
    \vspace{-2mm}
    \caption{(Colour online) The same as in figure \ref{fig:prob_classXY}, but for the PF model. In this case, the critical temperature is affected by the width of the range training temperatures
    and a relatively wide range is necessary to obtain precise $T_{\rm BKT}$ [c.f. figure~\ref{fig:prob_PF}~(b)]. The temperatures at which the ANN was trained 
    and tested were (in units of $t$): from 0.02 to 0.20 with stepsize 0.01.}
    \label{fig:prob_PF}
\end{figure}

In the case of the PF model, the results are different. As can be seen in figure 
\ref{fig:prob_PF}, the spread of the calculated probabilities is less dependent on $m$,
but the average critical temperature strongly depends on~$m$. This means that for the PF
model, an increase of the number of configurations used at the stage of phase classification
will not guarantee a more precise estimation of $T_{\rm BKT}$. Instead, for this model,
a sufficiently wide range of temperatures at which configurations for training are 
generated is necessary. From the physical point of view, this means that in the PF model,
extremely low-temperature configurations and extremely high-temperature configurations
are more different from the configurations close to the critical point than in the case
of the c-XY model. It can be seen in figure~\ref{fig:prob_PF}~(d) that for the PF model, the
relative error for $m=1$ extends to more than 20\%.

\begin{figure}[!t]
    \centering
    \includegraphics[width=0.63\textwidth]{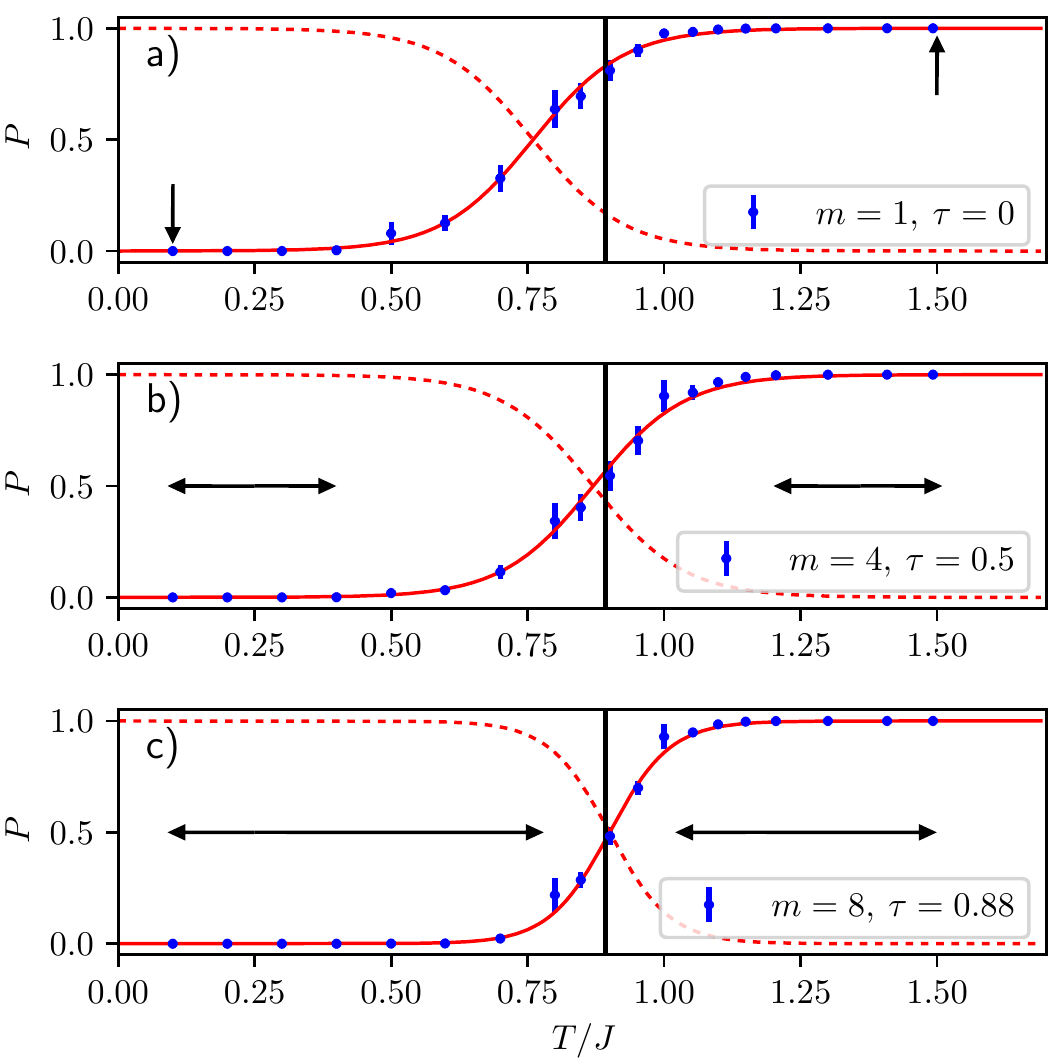}\raisebox{.3\height}{\includegraphics[width=0.35\textwidth]{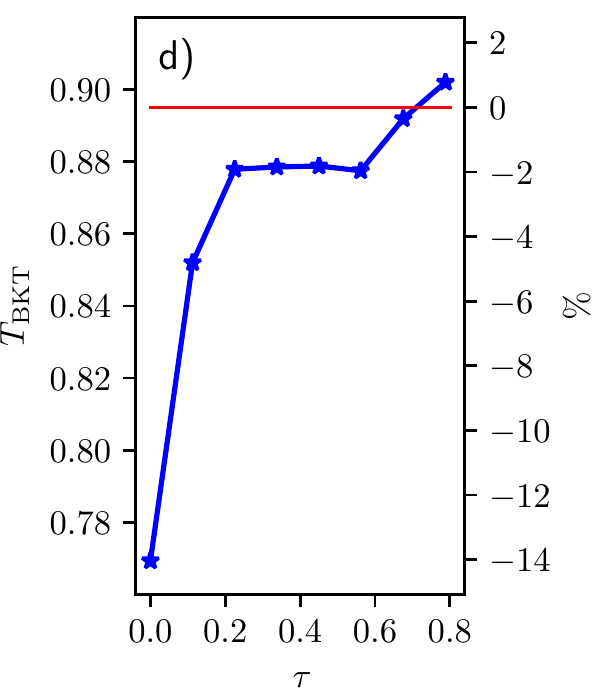}}
    \vspace{-2mm}
    \caption{(Colour online) The same as in figure \ref{fig:prob_classXY}, but for the q-XY model. Similarly to the c-XY model, also here the critical temperature is rather insensitive to the 
    width of the training temperatures, but the spread of the results decreases with an increasing $m$. The temperatures at which the ANN was trained 
    and tested were (in units of $J$): from 0.1 to 0.8 and from 1.2 to 1.5 with stepsize 0.1
    and from 0.85 to 1.15 with stepsize 0.05.}    
    \label{fig:prob_quantXY}
\end{figure}

The distribution of probabilities calculated for the q-XY model is similar to that for 
its classical counterpart. It is presented in figure \ref{fig:prob_quantXY}. Though for 
$m=1$, the estimated critical temperature differs from 
its real value by 14\% [see figure~\ref{fig:prob_quantXY}~(d)], the difference decreases very 
quickly with an increasing $m$ and already for $m\geqslant 3$, the relative error is around 2\%. 

The results show that in the case of the PF model, a much richer set of configurations
is required to properly train the ANN than for the XY models. The reason can be connected 
to a different 
character of this model. In both the classical and quantum XY models, the interaction 
range is limited to nearest neighbors. On the other hand, fermions in the PF model
mediate the effective interactions between arbitrarily spaced lattice sites. This
effect is seen in figure \ref{fig:hel_mod}: at low temperature the helicity modulus in the
c-XY and q-XY models converges even for very small systems. This is not the case for
the PF model, where even at very low temperatures (i.e., in an almost fully ordered 
state) the energy per lattice site depends on the system size. This results from the
delocalization of fermions which in the ordered state are similar to quantum particles in 
an infinite quantum well, with their energies strongly dependent on the size of the well.

\section{Summary}

We have demonstrated how the accuracy of finding the BKT transition in three different models
depends on the range of temperatures at which the ANN was trained. We used a simple 
feedforward network with densely connected hidden layers. We did not perform any feature 
engineering of the spin configurations generated in MC simulations and we did not use
convolutional layers. Therefore, the phase classification was based on raw spin configurations, 
not on the explicitly extracted vortices as in  \cite{melko2018}. Nevertheless, in figure~\ref{fig:prob_classXY}~(c) 
we compare the calculated probabilities and magnetization that results
from the finite size of the system. One can see there that the section of $P(T)$ that indicates
the BKT transition is much steeper than the temperature dependence of the magnetization, even 
if the network was trained on the extreme temperatures [figure~\ref{fig:prob_classXY}~(a)].
Therefore, we believe that the ANN learns not only the magnetization (which would vanish
in the thermodynamic limit), but also some topological features connected with the BKT
transition. One also cannot exclude that the ANN is capable of learning the character of the
spin-spin correlations which change their behavior at the BKT transition.
To confirm this, however, at least a finite size analysis of the ML results
would be necessary, which has not been performed here. However, our aim was different --- we wanted
to demonstrate how the
critical temperature determined by the ANN depends on the composition of the training set.
As one can expect, the larger is the variety of the configurations representing the low-temperature
and high-temperature phases, the better is the accuracy of the critical temperature. We found,
however, that for the c-XY and q-XY model, the average $T_{\rm BKT}$ was close to the actual 
one even if the ANN was trained relatively far from the critical point. Increasing the range of 
temperatures at which the network was trained, only slightly improves the numerical accuracy
(i.e., the difference between the average $T_{\rm BKT}$ determined by the ANN and the value
found from the RG equations),
but significantly reduces the uncertainty. The situation is different for the PF model, where 
the numerical accuracy is strongly dependent on the temperature range used at the training 
stage. We attribute this behavior to the long-range effective interaction present in the PF
model which can lead to a longer range of the spin-spin correlations and their different 
temperature dependence. 

Despite the difference in the results for the XY models and the PF model, in all cases it
is important to train the ANN not only at very low and at very high temperatures, but also 
as close as possible to the critical temperature. The main problem in training at temperatures 
close to $T_{\rm BKT}$ is that for supervised learning, the configurations must be labelled, so 
one should know at least an approximate value of the critical temperature. One of the ways to 
overcome this difficulty is to use the {\em learning by confusion} approach \cite{phtrans7} 
based on a combination of supervised and unsupervised techniques.

\section*{Acknowledgements}
M.M.M. acknowledges support by NCN (Poland) under grant 2016/23/B/ST3/00647. H.K. and N.T. acknowledge funding from grant no. NSF DMR 1629382.

\newpage

 \ukrainianpart
 
 \title{Застосування машинного навчання до переходу
 Березинського-Костерліца-Таулесса в класичних і квантових моделях}
 \author{М. Ріхтер-Лясковська\refaddr{US}, Г. Хан\refaddr{OH}, Н.
 Тріведі\refaddr{OH}, М.М. Маська\refaddr{US}}
 \addresses{
     \addr{US}Інститут фізики, Сілезький університет, вул. 75-го піхотного полку,
 1, 41-500 Хожів, Польща
     \addr{OH}Факультет фізики, університет штату Огайо, просп. В. Вудраффа, 191,
 Колумбус, Огайо 43210, США
 }
 
 \makeukrtitle
 
 \begin{abstract}
     \tolerance=3000%
     Перехід Березинського-Костерліца-Таулесса є дуже специфічним фазовим
 переходом, при якому всі термодинамічні величини є неперервними. Тому важко
 точно визначити критичну температуру. У цій статті нами показано, як можна
 використати нейронні мережі для розв'язання цього завдання. Зокрема,
 досліджено, до якої міри точність розпізнавання переходу залежить від способу
 навчання нейронних мереж. Ми застосовуємо наш підхід до трьох різних систем:
 (i) класична XY модель, (ii) фазово-ферміонна модель із взаємодією між
 класичними й квантовими ступенями вільності та (iii) квантова XY модель.
     \keywords фазові переходи, топологічні дефекти, XY модель, штучні нейронні
 мережі, машинне навчання
     
 \end{abstract} 

\begin{thebibliography}{99}
\bibitem{BKT1} Berezinskii V.L., Sov. Phys. JETP, 1971, \textbf{32}, 493.
\bibitem{BKT1a} Berezinskii V.L., Sov. Phys. JETP, 1972, \textbf{34}, 610.
\bibitem{BKT2} Kosterlitz J.M., Thouless D.J.,  J. Phys. C: Solid State Phys., 1972, \textbf{5}, L124, \doi{10.1088/0022-3719/5/11/002}.
\bibitem{BKT2a} Kosterlitz J.M., Thouless D.J., J. Phys. C: Solid State Phys., 1973, \textbf{6}, 1181, \doi{10.1088/0022-3719/6/7/010}.
\bibitem{phtrans} Carrasquilla J., Melko R.G., Nat. Phys., 2017, \textbf{13}, 431, \doi{10.1038/nphys4035}.
\bibitem{ising} Morningstar A., Melko R., J. Mach. Learn. Res., 2018, \textbf{18}, 163.
\bibitem{ising1} Ponte P., Melko R., Phys. Rev. B, 2017, \textbf{96}, 205146, \doi{10.1103/PhysRevB.96.205146}.
\bibitem{phtrans1} Zhang Y., Kim E.-A., Phys. Rev. Lett., 2017, \textbf{118}, 216401, \doi{10.1103/PhysRevLett.118.216401}.
\bibitem{phtrans2} Wang L., Phys.  Rev. B, 2016, \textbf{94}, 195105, \doi{10.1103/PhysRevB.94.195105}.
\bibitem{phtrans3} Hu W., Singh R.R.P., Scalettar R.T., Phys. Rev. E, 2017, \textbf{95}, 062122, \doi{10.1103/PhysRevE.95.062122}. 
\bibitem{phtrans4} Wetzel S.J., Phys. Rev. E, 2017, \textbf{96}, 022140, \doi{10.1103/PhysRevE.96.022140}.
\bibitem{phtrans5} Broecker P., Carrasquilla J., Melko R.G., Trebst S., Sci. Rep., 2017, \textbf{7},
8823,\\ \doi{10.1038/s41598-017-09098-0}.
\bibitem{phtrans6} Ch'ng K., Carrasquilla J., Melko R.G., Khatami E., Phys. Rev. X, 2017, \textbf{7}, 031038,\\ \doi{10.1103/PhysRevX.7.031038}.
\bibitem{phtrans7} Van Nieuwenburg E.P.L., Liu Y.-H., Huber S.D., Nat. Phys., 2017, \textbf{13}, 435, \doi{10.1038/nphys4037}. 
\bibitem{qml} Torlai G., Mazzola G., Carrasquilla J., Troyer M., Melko R., Carleo G., Nat. Phys., 2018, \textbf{14}, 447, \doi{10.1038/s41567-018-0048-5}.
\bibitem{qml1} Carleo G., Troyer M., Science, 2017, \textbf{355}, 602, \doi{10.1126/science.aag2302}.
\bibitem{melko2018} Beach M.J.S., Golubeva A., Melko R.G., Phys. Rev. B, 2018, \textbf{97}, 045207, \doi{10.1103/PhysRevB.97.045207}.
\bibitem{sarma} Deng D.-L., Li X., Sarma S.D., Phys. Rev. B, 2017, \textbf{96}, 195145, \doi{10.1103/PhysRevB.96.195145}.
\bibitem{zhang2018} Zhang W., Liu J., Wei T.-C., Preprint \arxiv{1804.02709}, 2018.
\bibitem{Nieva} Rodriguez-Nieva J.F., Scheurer M.S., Preprint \arxiv{1805.05961}, 2018.
\bibitem{pf} Ma\'ska M.M., Trivedi N., Preprint \arxiv{1706.04197}, 2017.
\bibitem{sandvik} Hsieh Y.-D., Kao Y.-J., Sandvik A.W., J. Stat. Mech.: Theory Exp., 2013, \textbf{2013}, P09001,\\ \doi{10.1088/1742-5468/2013/09/P09001}.
\bibitem{bf} Micnas R., Ranninger J., Robaszkiewicz S., Rev. Mod. Phys., 1990, \textbf{62}, 113, \doi{10.1103/RevModPhys.62.113}.
\bibitem{FK} Ma\'ska M.M., Czajka K., Phys. Rev. B, 2006, \textbf{74}, 035109, \doi{10.1103/PhysRevB.74.035109}.
\bibitem{helicity} Fisher M.E., Barber M.N., Jasnow D., Phys. Rev. A, 1973, \textbf{8}, 1111, \doi{10.1103/PhysRevA.8.1111}.
\bibitem{KRG} Kosterlitz J.M., J. Phys. C: Solid State Phys., 1974, \textbf{7}, 1046, \doi{10.1088/0022-3719/7/6/005}.
\bibitem{Schultka} Schultka N., Manousakis E., Phys. Rev. B, 1994, \textbf{49}, 12071,
\doi{10.1103/PhysRevB.49.12071}.
\bibitem{Tomita} Tomita Y., Okabe Y., Phys. Rev. B, 2002, \textbf{65}, 184405, \doi{10.1103/PhysRevB.65.184405}.
\bibitem{Weber} Weber H., Minnhagen P., Phys. Rev. B, 1988, \textbf{37}, 5986(R), \doi{10.1103/PhysRevB.37.5986}.
\bibitem{initialization} Sussillo D., Abbott L.F., Preprint \arxiv{1412.6558v3}, 2015.
\bibitem{tikhonov} Ng A.Y., In: Proceedings of the Twenty-First International Conference on Machine Learning (Banff, Canada, 2004), ACM, New York, 2004, 78, \doi{10.1145/1015330.1015435}.
\bibitem{cross-val} Gunasegaran T., Cheah Y.-N., In: Proceedings of the 8th International Conference on Information Technology (Amman, Jordan, 2017), IEEE, 2017, 89--95, \doi{10.1109/ICITECH.2017.8079960}.
\end{thebibliography}
\end{document}